\documentstyle[preprint,aps]{revtex}
\newcommand{\be}{\begin{equation}} \newcommand{\ee}{\end{equation}} 
\newcommand{\bea}{\begin{eqnarray}}\newcommand{\eea}{\end{eqnarray}}

\begin{document}
\draft
\preprint{OCHA-SP-00-06, hep-th/0012058}
\title{ Spin chains from super-models}
\author{Tetsuo Deguchi$^{*}$ and Pijush K. Ghosh$^{**}$}
\address{ 
Department of Physics,
Ochanomizu University,\\
2-1-1 Ohtsuka, Bunkyo-ku,
Tokyo 112-8610.\\}
\footnotetext{$\mbox{}^*$E-mail address: 
deguchi@phys.ocha.ac.jp}  
\footnotetext{$\mbox{}^{**}$E-mail address: 
pijush@degway.phys.ocha.ac.jp}  
\maketitle
\begin{abstract} 
We construct and study a class of $N$ particle supersymmetric Hamiltonians with
nearest and next-nearest neighbor inverse-square interaction in one dimension.
We show that inhomogeneous XY models in an external non-uniform
magnetic field can be obtained from these super-Hamiltonians in a particular
limit decoupling the fermionic degrees of freedom from the kinematic ones.
We further consider a suitable deformation of these super-models such that
inhomogeneous XXZ Hamiltonians in an external non-uniform magnetic
field are obtained in the same limit. We show that this
deformed Hamiltonian with rational potential is, (i) mapped to a
set of free super-oscillators through a similarity transformation
and (ii) supersymmetric in terms of a new, non-standard realization of
the supercharge. We construct many exact eigenstates of this
Hamiltonian and discuss about the applicability of this technique to
other models.
\end{abstract}

\section{Introduction}

The Calogero-Moser-Sutherland (CMS) system is a class of exactly solvable
models in one dimension with each particle interacting with the rest of the
particles through a long range inverse-square interaction
\cite{cs,cs1,pr,fl,bh,hf,pat,pani,me,fm,me1,ms,hs,gt}. These models
have been studied over a period of more than thirty years and have been found
to be related, directly or indirectly, to a wide class of physical
and mathematical systems starting from condensed matter physics to black holes.
One such interesting connection is that the CMS Hamiltonians are
related to Random Matrix Theory (RMT) describing Brownian motion model of
Dyson\cite{cs1,dy}. Since the RMT also plays an important role in describing the
level statistics of chaotic systems\cite{ls}, this connection shows an
universal aspect of diverse branches of physics\cite{als}.

Recently, a class of new models with long-range inverse-square
interaction among the particles has been introduced\cite{kha}. However,
unlike in CMS, particles in these models interact only with the
nearest-neighbor(NN) and the next-nearest-neighbor(NNN). Although one would
expect each particle to interact with the rest of the particles for systems
with long-range interaction, the motivation for considering such an NN-
variant of the CMS model stems from its close connection with Random Banded
Matrix (RBM) theory describing short-range Dyson model\cite{rbm}.
In particular, the norm of the
ground-state of this NN-variant of CMS system can be identified with the
joint-probability distribution function of the Gaussian RBM theory. 
As a result, different correlation functions can be calculated exactly
from the known results of RBM theory. The short-range Dyson model was
introduced  to understand the level statistics of pseudo-integrable
systems\cite{pint} like Aharanov-Bohm billiards\cite{aabb}, three dimensional
Anderson model at the metal-insulator transition point\cite{ami} and polygonal
billiards\cite{rbm}. Thus, as in the case of CMS system, these models play an
important role in the universal framework of diverse branches of physics like
RBM, pseudo-integrable system and short-range Dyson model.

The purpose of this paper is to study the supersymmetric version of this
relatively new class of CMS-type models with NN and NNN interactions.
In particular, we first construct the supersymmetric version of these
models. Using the Jordan-Wigner transformation
we find that these super-Hamiltonians indeed describe a class of
inverse-square
Hamiltonians with internal spin-degrees of freedom. The spin part of the
Hamiltonian describes a NN XY model in an external magnetic field with the
hopping-coefficient proportional to the inverse-square of the distance
between the NN particles. We further show that there exists a natural
`Freezing Limit'(FL) for all of these models with stable classical equilibrium
configurations, in which the particle and
the spin-degrees of freedom decouple completely. Thus, exact eigenstates
of these XY models can be obtained from the corresponding exact states
of the super-model.

We show that the rational model of this NN variant of CMS system is mapped
to a set of free superoscillators through a similarity transformation. This is
true for both $A_{N+1}$ and $BC_{N+1}$ type rational super-models. Using
this mapping we are able to obtain many exact eigenstates of
the rational super-models from the free super-oscillator basis.
In absence of any such mapping, we are able to obtain
only the ground-state of the Trigonometric and the Hyperbolic super-models.

We suitably deform the super-model such that the XXZ Hamiltonian is contained
in the fermionic part of the model and spin-degrees of freedom can be decoupled
from the coordinate degrees one in the FL. We also make sure that the deformed
Hamiltonian can be shown to be mapped to free-super oscillators through
a similarity transformation for the rational case. The realization of the
supersymmetry in the deformed model in terms of the usual representation of
the super-charges is lost. However, in terms of a non-standard representation
of the supercharge, the supersymmetry is still present in the deformed-model
with rational potential.
We obtain the exact ground-state and the first excited state of this deformed
super-Hamiltonian and, hence of the corresponding XXZ model obtained in the FL.
Though we are not able to go beyond at this
point, it is expected that the relation between the super-model and the free
super-oscillator would be helpful in obtaining other excited states and may be
even the complete spectrum, in future.
We also find the exact groundstate of the deformed model corresponding to
the super-model of trigonometric and hyperbolic type.

The plan of the paper is the following. We first briefly review the
supersymmetric quantum mechanics with many degrees of freedom in the
next section. We then show in Sec. III that the most general
super-model with  stable classical equilibrium configurations
has a natural FL in which the particle degrees of
freedom completely decouple from the fermionic degrees of freedom. We
also determine restrictions on the superpotential such that the fermionic
part of a general super-model describes an XY model. In Sec. III.B, we
show that the superpotential of the NN-variant of CMS system satisfies
this property and study the corresponding rational super-model in some details.
We show the mapping of this model to free superoscillators
through a similarity transformation and construct many exact eigenstates.
A suitable deformation of the super-model to study the XXZ model in the
FL has been considered in Sec. IV. We show the mapping of this 
deformed Hamiltonian to free superoscillators and construct a few exact
eigenstates. Finally in Sec. V., we conclude about our findings and
discuss on possible future courses of study. The results concerning
Jordan-Wigner transformation are reproduced in Appendix A. In appendix
B, a class of quadratic superpotentials which give rise to super-Hamiltonians
with XY model in its fermionic part is discussed. The fermionic part
of the super-model automatically decouples from the parent model for such
choices of quadratic superpotentials. We notice that the supersymmetric
phase of such super-models corresponds with the ferromagnetic phase of
the corresponding XY models. In Appendix C,
we discuss about the classical minimum equilibrium configurations 
of the rational inverse-square models. The Trigonometric and the Hyperbolic
inverse-square super-models are introduced and studied in 
Appendices D and E, respectively. Finally, in Appendix F, we discuss
the $OSP(2|2)$ superalgebra of the general inverse-square rational
super-models.

\section{SUSY QM with many degrees of freedom: brief review }

The supercharge $Q$ and its conjugate $Q^{\dagger}$ are defined as,
\be
Q=\sum_{i=1}^N \psi_i^{\dagger} \ a_i, \ \ \ \ 
Q^\dagger = \sum_{i=1}^N \psi_i \  a_i^{\dagger},
\label{eq0}
\ee
\noindent where the fermionic variables $\psi_i$'s satisfy the Clifford algebra,
\be
\{\psi_i,\psi_j\}=0=\{\psi_i^{\dagger},\psi_j^{\dagger}\}, \ \
\{\psi_i, \psi_j^{\dagger}\}=\delta_{ij}, \ \ i, j=1, 2, \dots, N.
\label{eq1}
\ee
\noindent The operators $a_i(a_i^{\dagger})$'s are analogous to bosonic 
annihilation ( creation ) operators. They 
are defined in terms of the momentum operators
$p_i=-i \frac{\partial}{\partial x_i}$ and the superpotential
$W(x_1, x_2, \dots, x_N)$ as,
\be
a_i= p_i - i W_i, \ \ a_i^{\dagger}= p_i + i W_i, \ \ 
W_i = \frac{\partial W}{\partial x_i},
\label{eq2}
\ee
\noindent and satisfy the following commutation relations among themselves,
\be
[a_i,a_j]=0=[a_i^{\dagger}, a_j^{\dagger}], \ [a_i, a_j^{\dagger}]=
[a_j, a_i^{\dagger}]= 2 W_{ij},
\ \ W_{ij}= \frac{\partial^2 W}{\partial x_i \partial x_j}.
\label{eq3}
\ee
\noindent Note that, by construction,  $W_i$'s satisfy the so called
`zero-curvature
condition' $\partial_i W_j = \partial_j W_i$. Also, for translationally
invariant superpotentials, these $W_i$'s satisfy the `sum to zero' condition,
$\sum_i W_i=0$. These two properties are useful ingredients in studying the
usual CMS model.

The supersymmetric Hamiltonian is defined in terms of the supercharges as,
\bea
H & = & \frac{1}{2} \{Q,Q^{\dagger}\}\nonumber \\
&= & \frac{1}{4} \sum_i \{a_i,a_i^{\dagger}\} + \frac{1}{4}
\sum_{i,j} [a_i,a_j^{\dagger}] [\psi_i^\dagger, \psi_j].
\label{eq4}
\eea
\noindent The Hamiltonian commutes with both $Q$ and $Q^{\dagger}$. The ground
state of $H$ is annihilated by both $Q$ and $Q^{\dagger}$.
Thus, the ground states are given by,
\be
\phi_0 = e^{-W} |0>, \ \ \phi_N = e^{W} |\bar{0}>,
\label{eq5}
\ee
\noindent where the fermionic vacuum $|0>$ and its conjugate
$|\bar{0}>$ in the $2^N$ dimensional fermionic Fock space are defined as,
\be
\psi_i |0> =0 , \ \ \ \psi_i^{\dagger} |\bar{0}>=0.
\label{eq6}
\ee
\noindent The first equation of (\ref{eq6}) defines the zero-fermion sector,
while the second one defines the $N$ fermion sector. If either $\phi_0$
or $\phi_N$ is normalizable, the
supersymmetry is preserved with zero ground state energy. On the other
hand, the supersymmetry is broken if neither $\phi_0$ nor $\phi_N$ is
normalizable. The ground state energy in this case is positive-definite.

\section{XY models and SUSY QM with nearest-neighbor interaction}
\subsection{General Case}

The supersymmetric Hamiltonian (\ref{eq4}) can be written as,
\be
H = \frac{1}{2} \sum_i \left ( p_i^2 + W_i^2 \right ) + \frac{1}{2}
\sum_{i,j} W_{ij} \left ( \psi_i^{\dagger} \psi_j - \psi_i \psi_j^{\dagger}
\right ).
\label{eq6.1}
\ee
\noindent Equation (\ref{eq3}) has been used to write Eq.
(\ref{eq6.1}) in its present form. The term $W_{ij}$ is independent of
the bosonic coordinates for the choice of a purely quadratic superpotential,
$W=\sum_{i,j} x_i S_{ij} x_j$ with $S$ any real, symmetric matrix.
Thus, the bosonic and the fermionic Hamiltonians in (\ref{eq6.1})
decouple automatically from each other. This class of superpotentials
producing XY models in the fermionic part have been discussed in Appendix B.
For all other nontrivial choices of $W$,
however, the bosonic and the fermionic degrees of freedom are coupled
together. They can be decoupled only in certain special limits. We find
that the FL described in \cite{fl} is well suited for the
Hamiltonian (\ref{eq6.1}) to decouple the fermionic part from its parent
Hamiltonian. In particular, let us assume that $W$ is proportional to an overall
coupling constant $\lambda$. Then the coefficient of the bosonic potential
becomes $\lambda^2$, while it is $\lambda$ for the fermionic part of the
Hamiltonian. Now we take the strong interaction limit $\lambda \rightarrow
\infty$. This can be achieved by multiplying $H$ with $\lambda^{-2}$ and taking
$\lambda \rightarrow \infty$. The leading relevant term in this limit is,
\be
H \equiv \frac{1}{2 \lambda}
\sum_{i,j} W_{ij} \left ( \psi_i^{\dagger} \psi_j - \psi_i \psi_j^{\dagger}
\right ) + O \left ( \lambda^{-2} \right ),
\label{eq6.11}
\ee
\noindent where the bosonic coordinates in $W_{ij}$ take the value of
their classical minimum equilibrium configurations, $W_i=0$. Note that
a leading,
but non-dynamical and irrelevant term $\sum_i W_i^2$ has not been included
in (\ref{eq6.11}), which identically vanishes for the classical minimum
equilibrium configurations.

We would now like to see if the superpotential can be chosen such that the
fermionic part of $H$ can be identified with some spin chain model in the
FL. We choose the superpotential $W$ such that,
\be
W_{ij} = \delta_{ij} g_i (x_1, x_2, \dots, x_N) + \delta_{i,j+1}
h_{i-1} (x_1, x_2, \dots, x_N)
+ \delta_{i,j-1} h_i (x_1, x_2, \dots, x_N),
\label{eq6.2}
\ee
\noindent where $h_i$'s and $g_i$'s are arbitrary functions of the bosonic
coordinates. With this choice of superpotential, the Hamiltonian now reads,
\be
H = \frac{1}{2} \sum_i \left ( p_i^2 + W_i^2 \right ) +
\frac{1}{2} \sum_i \left [ g_i n_i + 2 h_i \left ( \psi_i^{\dagger} \psi_{i+1}
- \psi_i \psi_{i+1}^{\dagger} \right ) \right ], \ \
n_i = [\psi_i^{\dagger}, \psi_i].
\label{eq6.3}
\ee
\noindent Note that the $n_i$' s are proportional to the fermionic number
operators $\psi_i^{\dagger} \psi_i$ for
each fermionic degree of freedom $\psi_i$ and the second term inside the
square bracket with the coefficient $h_i$ is the XY term at site $i$ in the
fermionic representation. In particular, using the Jordan-Wigner transformation
given in Appendix A with the periodic boundary conditions, we have,
\be
H = \frac{1}{2} \sum_i \left ( p_i^2 + W_i^2 \right ) +
\frac{1}{2} \sum_i \left [ g_i \sigma_i^z + h_i \left ( 
\sigma_i^{x} \sigma_{i+1}^x + \sigma_i^y \sigma_{i+1}^y \right ) \right ].
\label{eq6.33}
\ee
\noindent Thus, the supersymmetric Hamiltonian $H$, in general,
describes an $N$ particle system with both kinematical and internal
spin degrees of freedom. Note that both $g_i$'s and $h_i$'s depend on
the bosonic coordinates.
In the FL, as described above, it is possible to decouple
the spin degrees of freedom from the coordinate degrees of freedom. For such
cases, solving the supersymmetric Hamiltonian $H$, one would in fact also be
able to solve the corresponding spin chain problem. We illustrate this point
with a few examples below. We use the Hamiltonian (\ref{eq6.3})
with the fermionic representation for the spin variables for all practical
purposes.

A comment is in order. The XY model as the fermionic part of a supersymmetric
quantum mechanical model has first been identified in Ref. \cite{ritten}. 
However, such identification for an arbitrary superpotential is not correct.
The Jordan-Wigner transformation maps a nearest-neighbor XY model to a
free fermionic theory with the hopping allowed to the next nearest site only
and the vice versa.
It is necessary that the superpotential has a form given by Eq. (\ref{eq6.2})
for such fermionic bilinear to be present in a supersymmetric theory.
Thus, for the first time in the literature, we give the necessary condition to
be satisfied by the superpotential so that nearest-neighbor XY models are
contained in the fermionic sector of the supersymmetric theory. Such a criteria
is missing in Ref. \cite{ritten} and, to the best of our knowledge,
in all subsequent papers on this subject. 
We would like to mention here that for a general superpotential which does not
satisfy (\ref{eq6.2}), one can still obtain a fermionic theory in the FL.
However, the Jordan-Wigner transformation may not be applicable to map such
a fermionic theory into a nearest-neighbor XY model.

\subsection{ Hamiltonians with Inverse-square interaction}

In this section, we consider rational super-models
with inverse-square interaction. The Trigonometric and the Hyperbolic
version of these classes of inverse-square models are discussed
in Appendix D and Appendix E, respectively. For the rational model,
we choose the super-potential of the following form,
\be
W = - w + \frac{\omega}{2} \sum_i x_i^2, \ \
w= ln \ G(x_1, x_2, \dots, x_N),
\label{sup}
\ee
\noindent where $G$ is a homogeneous function of any positive
degree $d$:
\be
\sum_i x_i \frac{\partial G}{\partial x_i} = d G.
\label{supsup}
\ee
\noindent The Hamiltonian for this choice of superpotential has a
dynamical $OSP(2|2)$ supersymmetry\cite{me}. The relevant algebra is given
in Appendix F. Using the subalgebra $O(2,1) \times U(1)$, one can show
that the Hamiltonian is mapped to a set of free superoscillators
through a similarity transformation\cite{me}. We use this result
very often in this section to construct exact eigenstates. We shall
specialize to two different choices of $G$.

A comment regarding the similarity transformation is in order. One might wonder
that any super-Hamiltonian with the superpotential $W$ described by
(\ref{sup}) and (\ref{supsup}) is exactly solvable, due to its mapping
to free super oscillators through similarity transformations. We would like
to point out that this may not be true always, because, merely mapping the
original Hamiltonian to free super-oscillators is not sufficient for such
conclusions. We have to make sure that
the similarity transformation, which is responsible for such mapping,
keeps the original Hamiltonian in its own Hilbert space. Thus, as a check,
one should show that the complete spectrum and the corresponding
well-behaved, normalizable eigen-functions of $H$ can be constructed from the
Hamiltonian of the free super-oscillators through inverse similarity
transformation. The formal equivalence at the operator level acts as a
necessary condition, while the construction of the complete spectrum and
associated well-behaved eigen-functions of the original Hamiltonian from the
super-oscillator model is sufficient to claim the equivalence between
these two Hamiltonians.

\subsubsection{Rational model of $A_{N+1}$-type}

We choose the superpotential as,
\be
G=\prod_{i=1}^N (x_i - x_{i+1})^{\lambda}, 
 \ \ x_{N+i}= x_i.
\label{eq16}
\ee
\noindent Note that $G$ is a homogeneous function of degree $d=\lambda N$.
This produces the following expressions for $g_i$ and $h_i$,
\be
h_i = - \lambda (x_i - x_{i+1})^{-2}, \ \
g_i = \omega  - (h_i + h_{i-1}).
\label{eq16.1}
\ee
\noindent The Hamiltonian $H$ for the rational superpotential now reads,
\bea
H_{R} & = & - \frac{1}{2} \sum_i \frac{\partial^2}{\partial x_i^2} +
\frac{\lambda^2}{2} \sum_i \left [ 2 ( x_i- x_{i+1})^{-2} 
- (x_{i-1}-x_i)^{-1} (x_i - x_{i+1})^{-1} \right ] 
 +  \frac{1}{2} \omega^2 \sum_i x_i^2\nonumber \\
&  - & \lambda \omega N
+ \lambda \sum_i \left [ ( x_i - x_{i+1} )^{-2} \left (
\frac{1}{2} ( n_i + n_{i+1} )
-\psi_i^{\dagger} \psi_{i+1} + \psi_i \psi_{i+1}^{\dagger} \right )
+ \frac{\omega}{2 \lambda} n_i \right ],
\label{eq17}
\eea
\noindent with the periodic boundary conditions on the fermionic variables:
$\psi_{N+i}= \psi_i$.
This is the supersymmetric generalization of a class of Hamiltonian
recently introduced and studied in \cite{kha}. Unlike in CMS Hamiltonians
\cite{cs,cs1,pr,bh}, particles in this model interact with each other
through the NN and the NNN interaction.
Note that the bosonic many-body interaction is of NN as well as NNN-type.
On the other hand, the fermionic interaction is only of NN-type. For the
special case of $N=3$,
$H_R$ reduces to the rational CMS super-Hamiltonian due to the periodic
boundary conditions imposed on the bosonic and the fermionic coordinates.
We would like to mention here that our whole analysis goes through also for
the open boundary conditions. The third
term with the coefficient $\lambda$ contains the XY Hamiltonian in terms of
fermionic variables. Thus, particles in this model are also having internal
spin degrees of freedom.

The groundstate of $H$ in the supersymmetric phase ( $\lambda > 0 $ ) is
given by,
\be
\phi_0 = \prod_i \left ( x_i - x_{i+1} \right )^{\lambda} 
e^{-\frac{\omega}{2} \sum_i x_i^2} \ |0>.
\label{eq17.0}
\ee
\noindent The wavefunction $\phi_0$ is normalizable for $\lambda >
- \frac{1}{2}$. However, a stronger criteria that each momentum operator
$p_i$ is self-adjoint for the wavefunction of the form $\phi_0$ 
requires $\lambda > 0$. The supersymmetry is preserved for $\lambda >0$,
while it is broken for $\lambda <0$. The groundstate of $H$ in the
supersymmetry-breaking phase ( $\lambda < 0$ ) is given by,
\be
\tilde{\phi}_0 = \prod_i \left ( x_i - x_{i+1} \right )^{- \lambda}
e^{-\frac{\omega}{2} \sum_i x_i^2} \ |\bar{0}>.
\label{eq17.00}
\ee
\noindent The wavefunction $\tilde{\phi}_0$ is normalizable for
$\lambda < \frac{1}{2}$. However, the self-adjointness criteria of each
momentum operator $p_i$ determines $\lambda < 0$. The eigenspectrum
of $H$ in the supersymmetry-breaking phase can be described by using
a `duality property' of this kind of Hamiltonian\cite{me}. Thus, we
consider only the supersymmetric phase of $H$ in this paper.
Note that the effect of going from the supersymmetric phase to
supersymmetry-breaking phase of $H$ is just an overall change in the sign
of the coupling constant for the corresponding spin chain model.
 
The total fermion number operator $N_f=\sum_i \psi_i^{\dagger} \psi_i$
commutes with the Hamiltonian
$H_R$ and $G$ is a homogeneous function of a fixed degree. Following the
treatment of Ref. \cite{me}, it can be shown that the Hamiltonian is mapped
to free superoscillators through a similarity transformation. In particular,
\bea
H_{sho} & = & T H_R T^{-1}\nonumber \\
&=& \omega \sum_i \left ( x_i \frac{\partial}{\partial x_i} + \psi_i^{\dagger}
\psi_i \right ),
\label{eq17.01}
\eea
\noindent where the operator $ T = e^{\frac{S}{2}} e^W$ is given by,
\bea
S & = & \frac{1}{2} \sum_i \frac{\partial^2}{\partial x_i^2} +
\lambda \sum_i (x_i - x_{i+1})^{-1} \left ( \frac{\partial}{\partial x_i}
- \frac{\partial}{\partial x_{i+1}} \right )\nonumber \\
& - & \lambda \sum_i (x_i - x_{i+1})^{-2} \left [ \left ( \psi_i^{\dagger}
\psi_i+ \psi_{i+1}^{\dagger} \psi_{i+1} \right )
- \left ( \psi_i^{\dagger}
\psi_{i+1} - \psi_i \psi_{i+1}^{\dagger} \right ) \right ].
\label{eq17.02}
\eea
\noindent The relation (\ref{eq17.01}) opens up the possibility of
constructing the eigenstates of $H_R$ from that of $H_{sho}$.
In particular, if $P_{n} |0>$ is an eigenfunction of $H_{sho}$
with the eigenvalue $E_{n}$, then, $H_R$ has the same eigenvalue
$E_{n}$ with the eigenfunction given by,
\be
\chi = T^{-1} P_{n} |0>.
\label{eq17.03}
\ee
\noindent $P_{n}$ in (\ref{eq17.03}), in general, is a function of both
the bosonic and the fermionic coordinates. If we choose $P_{n}$ to
be a single-particle wave-function of $H_{sho}$, $\chi$ contains an essential
singularity. This is not physically acceptable. This shows that the original
many-body problem is a highly correlated system and we have to choose $P_{n}$
as suitable combinations of all the single-particle states. We find that the
following choices of $P_{n}$,
\bea
P_{n}^I & = & \sum_i x_i^n \psi_i^{\dagger},\ \
P_{n}^{II} = \sum_{i \neq j} ( x_i^n - x_j^n) \psi_i^{\dagger}
\psi_j^{\dagger},\ \ n=0,1,2, 3,\nonumber \\
P^{III} & = & \sum_{i \neq j \neq k} Det \left ( \matrix{ {1} & {1}
& {1} \cr {x_i^m} & {x_j^m} & {x_k^m} \cr
{x_i^{n}} & {x_j^{n}} & {x_k^{n}} } \right )
\psi_i^{\dagger} \psi_j^{\dagger} \psi_k^{\dagger}, \ \
m \neq n , \ \ m, n=1, 2, 3,\nonumber \\
P^{IV} & = & \sum_{i \neq j \neq k \neq l} Det \left ( \matrix{ {1} & {1}
& {1} & {1} \cr {x_i} & {x_j} & {x_k} & {x_l} \cr
{x_i^2} & {x_j^2} & {x_k^2} & {x_l^2} \cr
{x_i^3} & {x_j^3} & {x_k^3} & {x_l^3} } \right ) \psi_i^{\dagger}
\psi_j^{\dagger} \psi_k^{\dagger} \psi_l^{\dagger},
\label{eq17.04}
\eea
\noindent result in well-behaved, non-singular $\chi$. Unlike in the case
of super-CMS system, there are no compelling reasons to consider only those
$P_{n}$ which are symmetric under the combined exchange of the bosonic and
the fermionic coordinates. Surprisingly, it turns out that the $P_{n}$
corresponding to the exact eigenstates $\chi$ found by us necessarily have this
property. This can be checked easily for $P_n^I$ and $P_n^{II}$. It may be
worth mentioning here that the matrices appearing in
$P_{III}$ and $P_{IV}$ above are the Slater determinants constructed out of
the single-particle bosonic wave-functions. As a result, they are antisymmetric
under the exchange of any two indices. Together with the antisymmetric nature
of the fermionic variables, it is easy to see that $P_{III}$ and $P_{IV}$
are invariant under the combined exchange of the bosonic and the
fermionic coordinates. Note that unlike in the rational super-CMS system,
the maximum power which a single-particle coordinate $x_i$ can have in $P_n^I$,
$P_n^{II}$, $P^{III}$ and $P^{IV}$, is required to be 3. The construction of
$P^k$ as a
Slater determinant in terms of single-particle bosonic states for
$ k \geq 5$ necessarily involves a single-particle coordinate with power
$n \geq 4$. Thus, we find all $P^k$ for $k \geq 5$ produces singular $\chi$
for the model under consideration. It may be recalled here that the
similar construction of $P^k$ with $k$ ranging from $1$ to $N$
produces well-behaved $\chi$ for the rational super-CMS system.

In equation (\ref{eq17.04}), the $P$'s correspond to `basic
excitations' with $N_f=1,2,3$ and $4$. One can construct many other
$P$ producing non-singular $\chi$ by suitably
combining these four $P$'s and the symmetric polynomials,
\be
r_{n,m} = (\sum_i x_i^n )^m, \ \ n=1, 2, 3; \ m=1, 2, \dots, \infty.
\ee
\noindent For example, the choice $P = r_{n,m} P_k^I P_l^{II}$ produces
non-singular $\chi$. We now give an algebraic construction of the spectrum.
Consider the following set of operators,
\bea
&& b_i^{-} =  i p_i = \frac{\partial}{\partial x_i}, \ \ \
b_i^{+}= 2 \omega x_i\nonumber \\
&& B_n^{-} = \sum_{i=1}^N T^{-1} b_i^{-^n} T, \ \ \
B_n^{+}=  \sum_{i=1}^N T^{-1} b_i^{{+}^n} T, \ \
n=1, 2, 3,\nonumber \\
&& F_m^{-}= \sum_i T^{-1} \psi_i b_i^{-^{m-1}} T,\ \ \
F_m^{+}=\sum_i T^{-1} \psi_i^{\dagger} b_i^{{+}^{m-1}} T, \ \
m=1, 2, 3, 4,\nonumber \\
&& q_n^{-}=T^{-1} \sum_i \psi_i^{\dagger} b_i^{-^n} T, \ \ \
q_n^{+}= T^{-1} \sum_i \psi_i b_i^{{+}^n} T, \ \
n=1, 2, 3.
\label{eq610}
\eea
\noindent Note that there are three bosonic anhilation(creation) operators
$B_n^{-}(B_n^{+})$ and four fermionic anhilation(creation) operators
$F_n^{-}(F_n^{+})$. This is because the highest power of $b_i^+$ should
be $3$. These operators satisfy the following algebra,
\bea
&& \{F_m^{+}, F_n^{+}\}=0, \ \ [B_m^{+},F_n^{+}]=0, \ \
[B_m^{+}, B_n^{+}]=0,\nonumber \\
&& \{q_1^{-}, F_n^{+}\}=0, \ \
\{q_1^{+}, F_n^{+}\}=B_n^{+}, \ \
[H, F_n^{+}]=n \omega F_n^{+},\nonumber \\ \ \
&& [q_1^{-},B_n^{+}]= 2 n \omega F_n^{+}, \ \ [q_1^{+}, B_n^{+}]=0, \ \
[H, B_n^{+}]= n \omega B_n^{+}.
\label{eq611}
\eea
\noindent The eigenstates can now be created in an algebraic manner by
using the above relations. In particular,
\be
\chi_{n_1 \dots n_{3} \nu_1 \dots \nu_4}= \prod_{k=1}^3 B_k^{+^{n_k}}
F_k^{+^{\nu_k}} F_4^{+^{\nu_4}} \ \phi_0,
\label{eq6111}
\ee
\noindent is the eigenfunction with the eigen-value,
\be
 E= \omega \left (\sum_{k=1}^{3} k (n_k + \nu_k) + 4 \nu_4 \right ).
\ee
\noindent  The bosonic quantum numbers $n_k$'s are
nonnegative integers, while the fermionic quantum numbers $\nu_k$'s are either
$0$ or $1$. Note that a system consisting of a set of three independent
super-oscillators with the frequencies $\omega, 2 \omega, 3 \omega$ and a
fermionic oscillator with the frequency $4 \omega$ have the same energy $E$.
Thus, the spectrum of the model
is identical to that of a system consisting of three independent
super-oscillators with the frequencies $\omega, 2 \omega, 3 \omega$ and a
fermionic oscillator with the frequency $4 \omega$.
The physical reason behind the disparity between the bosonic and the fermionic
excitations deserves further understandings.

In order to take the FL described in \cite{fl}, we first scale $\omega$ as
$\lambda \omega$ and then take the limit $\lambda \rightarrow \infty$.
The spin degrees of freedom decouple completely from the
kinematical ones in this strong interaction limit.
The dynamics of the system is lost and the particles are frozen at their
classical equilibrium configurations,
\be
W_i = \lambda \omega x_i + \lambda \left [ (x_{i-1} - x_i )^{-1} -
(x_i - x_{i+1})^{-1} \right ] = 0.
\label{eq17.1}
\ee
\noindent In particular, we multiply the Hamiltonian $H$ with $\lambda^{-2}$
and take the limit $\lambda \rightarrow \infty$. The relevant leading term in
this limit gives an XY Hamiltonian,
\be
H_s = \sum_i \left [ ( x_i - x_{i+1} )^{-2} \left (
\frac{1}{2} ( n_i + n_{i+1} )
-\psi_i^{\dagger} \psi_{i+1} + \psi_i \psi_{i+1}^{\dagger} \right )
+ \frac{\omega}{2} n_i \right ],
\label{eq18}
\ee
\noindent where $x_i's$ are determined from (\ref{eq17.1}). We are not able to
solve (\ref{eq17.1}) for arbitrary $N$. However, we show in Appendix C that 
the equilibrium configurations of the particles necessarily
constitute a non-uniform lattice.

\subsubsection{Rational model of $BC_{N+1}$-type }

We choose the superpotential as,
\be
G(\lambda, \lambda_1, \lambda_2) = \prod_i (x_i^2 - x_{i+1}^2)^{\lambda}
\prod_j x_j^{\lambda_1} \prod_k (2 x_k)^{\lambda_2},
\label{gg}
\ee
\noindent where $\lambda$, $\lambda_1$ and $\lambda_2$ are arbitrary
parameters. The $D_{N+1}$-type model is described by taking 
$\lambda_1=\lambda_2=0$,
while the case $\lambda_1=0$($\lambda_2=0$) describes $C_{N+1}$($B_{N+1}$)-type
model. We restrict ourselves to the $B_{N+1}$-type Hamiltonian, 
without loss of any generality. We introduce two new variables as,
\be
q_i = x_i - x_{i+1}, \ \ \bar{q}_i = x_i + x_{i+1}.
\ee
\noindent The Hamiltonian $H$ in
terms of these new variables is given by,
\bea
H & = & \frac{1}{2} \sum_i p_i^2 + \lambda^2 \sum_i \left [
q_i^{-2}+ \bar{q}_i^{-2} - (q_{i-1}^{-1} - \bar{q}_{i-1}^{-1})
(q_i^{-1} + \bar{q}_i^{-1}) \right ] + \frac{1}{2} \sum_i \left ( 
\omega^2 x_i^2 + \frac{\lambda_1}{ x_i^2} \right )\nonumber \\
& - & \lambda \omega \left (2 + \frac{\lambda_1}{\lambda} \right) N
+ \frac{\lambda}{2} \sum_i q_i^{-2} 
\left [ (n_i + n_{i+1} )-
2 \left (\psi_i^{\dagger} \psi_{i+1} - \psi_i \psi_{i+1}^{\dagger} \right )
\right ]\nonumber \\
& + & \frac{\lambda}{2} \sum_i \bar{q}_i^{-2} 
\left [ (n_i + n_{i+1} ) + 2 \left (\psi_i^{\dagger} \psi_{i+1} -
\psi_i \psi_{i+1}^{\dagger} \right ) \right ] +
\sum_i \frac{1}{2} \left ( \omega +
\frac{\lambda_1}{x_i^2} \right ) n_i.
\label{bcn}
\eea
\noindent The ground-state in the supersymmetry-preserving phase
( $\lambda >0$, $\lambda_1 > 0$ ) is given by,
\be
\psi_0 = \prod_i (x_i^2 - x_{i+1}^2)^{\lambda}
\prod_j x_j^{\lambda_1} e^{- \frac{\omega}{2} \sum _i x_i^2} |0>.
\ee
\noindent On the other hand, in the supersymmetry breaking phase, the
groundstate has the following expressions,
\bea
&& \psi_0^I = \prod_i (x_i^2 - x_{i+1}^2)^{-\lambda} \prod_j x_j^{-\lambda_1}
e^{- \frac{\omega}{2} \sum _i x_i^2} |\bar{0}>, \ \
\lambda, \lambda_1 < 0,\nonumber \\
&& \psi_0^{II} = \prod_i (x_i^2 - x_{i+1}^2)^{-\lambda} 
\prod_j x_j^{1+\lambda_1} e^{- \frac{\omega}{2}
\sum _i x_i^2} |\bar{0}>, \ \lambda < 0, \lambda_1 > 0.
\eea
\noindent We do not know the groundstate in the region $\lambda > 0$,
$\lambda_1 < 0$. This is partially because the known method \cite{me}
uses the `shape-invariance' property of the many-body potential to
determine the groundstate in such supersymmetry breaking phases.
Unfortunately, the many-body potential in the present case is not
`shape invariant' due to the presence of the three-body term. It might be
recalled here that in the case of $BC_{N+1}$ type super-CMS, there is no
three-body term and the many-body potential is `shape invariant'\cite{me,me1}.
Thus, the groundstate of super-CMS of $BC_{N+1}$ type can be obtained in
all the regions of the parameter space.

We study only the supersymmetric phase of $H$ in this section and construct
some exact eigenstates. Following \cite{me}, similar construction of exact
eigenstates in the two supersymmetry-breaking phases associated
with the groundstates $\psi_0^I$ and $\psi_0^{II}$ can be carried over in a
straightforward way.

The Hamiltonian (\ref{bcn}) can be mapped to a set of
free superoscillators through a similarity transformation,
$H_{sho} = {\cal{T}} H {\cal{T}}^{-1}$, where the $W$ in
${\cal{T}}=e^{\frac{\hat{S}}{2}} e^W$ is
determined from (\ref{gg}) and $\hat{S}$ is given by, 
\bea
\hat{S} & = & \sum_i \left [- \frac{1}{2} p_i^2 + \lambda ( q_i^{-1} +
\bar{q}_i^{-1} + \frac{\lambda_1}{\lambda} x_i^{-1} )
\left ( \frac{\partial}{\partial x_i} -
\frac{\partial}{\partial x_{i+1}} \right ) \right ]
 - \lambda \sum_i q_i^{-2} (\psi_i^{\dagger}\psi_i + 
\psi_{i+1}^{\dagger} \psi_{i+1} )\nonumber \\
& + & \lambda \sum_i q_i^{-2}
\left (\psi_i^{\dagger} \psi_{i+1} - \psi_i \psi_{i+1}^{\dagger} \right )
+ \lambda \sum_i \bar{q}_i^{-2} 
\left [ (\psi_i^{\dagger} \psi_i + \psi_{i+1}^{\dagger}\psi_{i+1} ) +
\left (\psi_i^{\dagger} \psi_{i+1} -
\psi_i \psi_{i+1}^{\dagger} \right ) \right ]\nonumber \\
& + & \sum_i \frac{\lambda_1}{2} x_i^{-2} n_i.
\eea
\noindent We find that the physically acceptable wave-function $\chi
= {\cal{T}}^{-1} P |0>$ of $H$ can be obtained for the following choices
of $P$, 
\bea
&& P_n^I = \sum_i x_i^{2 n} (x_i \psi_i^{\dagger}), \ n=0, 1,\nonumber \\
&& P^{II} = \sum_{i \neq j} (x_i^2 - x_j^2) (x_i \psi_i^{\dagger})
(x_j \psi_j^{\dagger}).
\eea
\noindent Note that both $P^I$ and $P^{II}$ are invariant under the following
two transformations,
(i) $ (x_i, \psi_i^{\dagger}) \leftrightarrow (x_j, \psi_j^{\dagger} )$
and (ii) $ (x_i, \psi_i^{\dagger}) \rightarrow (-x_i, -\psi_i^{\dagger})$.
These are also the symmetry of the corresponding expression for
the $BC_{N+1}$-type rational super-CMS\cite{me}. We are unable to find any
other $P$ which produces nonsingular $\chi$. Note that the maximum power
of a single-particle coordinate is given by $3$ in both $P_I$ and $P_{II}$. Any
attempt to construct other $P$ with higher fermion numbers and having
the properties (i) and (ii) necessarily involves a single-particle
coordinate with power greater than three. This produces singular $\chi$.

The eigen-spectrum also can be constructed in an algebraic way. We define
the following set of creation and annihilation operators,
\be
{\cal{B}}_{1}^+ = {\cal{T}}^{-1} \sum_i b_i^{+^{2}} {\cal{T}}, \ \
{\cal{F}}_{n}^+ = {\cal{T}}^{-1} \sum_i \psi_i^{\dagger}
b_i^{+^{2 n-1}} {\cal{T}}, \ \ n=1, 2.
\label{eqq}
\ee
\noindent These operators satisfy an algebra similar to (\ref{eq611}).
Note there are only one bosonic operator, while two are fermionic
operators. This is again related to the fact that the highest power of
$b_i^{\dagger}$ should be 3. The eigen-states are obtained as,
\be
{\cal{\chi}}_{n_1, \nu_1, \nu_2}=
{\cal{B}}_1^{+^{n_1}} {\cal{F}}_1^{+^{\nu_1}} {\cal{F}}_2^{+^{\nu_2}}
\ \phi_0,
\label{eq66}
\ee
\noindent with the energy ${\cal{E}}= 2  \omega ( n_1 + \nu_1) + 4 \omega
\nu_2$. The bosonic quantum number is non-negative integers, while the fermionic
quantum numbers are $0$ or $1$. The spectrum is that of a superoscillator
with the frequency $2 \omega$ and a fermionic oscillator with the frequency
$4 \omega$.

\section{XXZ model}

It is known that the XXZ model describes an interacting fermionic theory
with the interaction term proportional to $ \sum_i n_i n_{i+1}$, which is
quartic in the fermionic variables. The super-Hamiltonian $H$ being
at most quadratic in these variables, our methods described in the previous
section should be modified to make room for the study of XXZ model.
We deform the supersymmetric Hamiltonian $H$ to ${\cal{H}}$ with the
deformation parameter $\delta$,
\bea
{\cal{H}} & = & H + 2 \delta \sum_i
 h_i \psi_i^{\dagger} \psi_i \psi_{i+1}^{\dagger} \psi_{i+1} \nonumber \\
& = & \frac{1}{2} \sum_i \left ( p_i^2 + W_i^2 \right ) + 
\frac{1}{2} \sum_i \left [ g_i n_i + 2 h_i \left ( \psi_i^{\dagger} \psi_{i+1}
- \psi_i \psi_{i+1}^{\dagger} + 2 \delta \psi_i^{\dagger} \psi_i
\psi_{i+1}^{\dagger} \psi_{i+1} \right ) \right ].
\label{x1}
\eea 
\noindent The XXZ Hamiltonian in an external magnetic field is contained
in ${\cal{H}}$,
\be
{\cal{H}} = \frac{1}{2} \sum_i \left (p_i^2 + W_i^2 \right )
 + \frac{1}{2} \sum_i \left [ g_i \sigma_i^z
+ h_i \left ( \sigma_i^x \sigma_{i+1}^x + \sigma_i^y \sigma_{i+1}^y
+ \delta (1+\sigma_i^z)(1+\sigma_{i+1}^z) \right ) \right ].
\ee
\noindent With the choice of a suitable $W$, the spin degrees of freedom
can be decoupled from the kinematical one in the FL. In the process, we
obtain an inhomogeneous XXZ model in an external magnetic field,
\be
H_{xxz} = \frac{1}{2} \sum_i \left [ \delta h_i +
(g_i + \delta h_i ) \sigma_i^z + \delta h_i \sigma_{i+1}^z
+ h_i \left ( \sigma_i^x \sigma_{i+1}^x + \sigma_i^y \sigma_{i+1}^y
+ \delta \sigma_i^z \sigma_{i+1}^z \right ) \right ], 
\label{x2}
\ee
\noindent with $g_i$ and $h_i$ determined by the classical minimum
equilibrium configurations of the Hamiltonian, $W_i=0$. Note that the
external magnetic field depends on the asymmetry parameter $\delta$. 
This will lead to important consequences on the phase structure of the spin
Hamiltonian. We would like to point out here that we have chosen a particular
deformation of $H$ for our discussions. One can choose to work with several
other deformations of the Hamiltonian $H$. A deformed-Hamiltonian different
from (\ref{x1}) has been presented in Appendix D describing Trigonometric model.
We use the fermionic representation (\ref{x1}) for our discussions
from now onwards.
  
Note that $\phi_0$ given in Eq. (\ref{eq5}) is an exact eigenstate of
${\cal{H}}$ with zero eigenvalue.
Now recall that $H$ is bounded from below owing to supersymmetry
and the fermionic quartic term in ${\cal{H}}$ is semi-positive. Thus,
this is indeed the exact groundstate of ${\cal{H}}$ with the constraints
$\delta h_i \geq 0$ for all $i$. The constraints for our choices of $W$
for the super-model discussed before imply a relation involving the
deformation-parameter $\delta$ and the many-body coupling of the super-model.
We now specialize to the rational $W$ only, since it has many advantages
over the Trigonometric and Hyperbolic models. The latter models are discussed
in Appendices D and E, respectively.

The supersymmetry is lost when $\delta \neq 0$, in terms of the usual
realization (\ref{eq0}) of the supercharge. However, the supersymmetry can
be realized for the rational super-model ${\cal{H}}$ in terms of some new,
non-standard representation of the supercharge.
We discuss here only the rational $W$ of $A_{N+1}$ type.
However, we should note that this is true for
any $W$ satisfying (\ref{sup}) and (\ref{supsup}).
Let us choose $g_i$ and $h_i$ such as given in
(\ref{eq16.1}). The Hamiltonian ${\cal{H}}_R$ which generalizes $H_R$
of (\ref{eq17}) now reads,
\bea
{\cal{H}}_R & = & - \frac{1}{2} \sum_i \frac{\partial^2}{\partial x_i^2} +
\frac{\lambda^2}{2} \sum_i \left [ 2 ( x_i- x_{i+1})^{-2}
- (x_{i-1}-x_i)^{-1} (x_i - x_{i+1})^{-1} \right ]
+ \frac{1}{2} \omega ^2 \sum_i x_i^2\nonumber \\
& + & \lambda \sum_i \left [ ( x_i - x_{i+1} )^{-2} \left (
\frac{1}{2} ( n_i + n_{i+1} )
-\psi_i^{\dagger} \psi_{i+1} + \psi_i \psi_{i+1}^{\dagger} - 2 \delta
\psi_i^{\dagger} \psi_i \psi_{i+1}^{\dagger} \psi_{i+1} \right )
+ \frac{\omega}{2 \lambda} n_i \right ]\nonumber \\
& - & \lambda \omega N .
\eea
\noindent Note that the wave-function given in (\ref{eq17.0}) still gives the
ground-state of ${\cal{H}}_R$ with zero energy for $\lambda > 0, \delta < 0$.
This is, in fact, the groundstate for arbitrary $\delta$. In order to see this,
we note that the bosonic potential appearing in the above Hamiltonian,
except for the confining harmonic term, is a homogeneous function of the
bosonic coordinates with degree $-2$. Following Ref. \cite{me}, the Hamiltonian
can be transformed to a free super-oscillators through a similarity
transformation,
\be
H_{sho} = T_1 {\cal{H}} T_1^{-1},
\ee
\noindent where $W$ in $T_1=e^{\frac{S_1}{2}} e^W$ is given by (\ref{eq16})
and $S_1$ is determined by,
\be
S_1 = S + 2 \delta \lambda \sum_i (x_i-x_{i+1})^{-2}
\psi_i^{\dagger} \psi_i \psi_{i+1}^{\dagger} \psi_{i+1}.
\ee
\noindent Here we recall that $S$ has been defined by (\ref{eq17.02}). Thus,
the deformed-Hamiltonian ${\cal{H}}_R$ is mapped to a
set of free superoscillators. Now we define the odd operator $q$ and its
conjugate $q^{\dagger}$ as,
\be
q= T_1^{-1} \sum_i \psi_i^{\dagger} b_i^- T_1, \ \
q^{\dagger} = T_1^{-1} \sum_i \psi_i b_i^+ T_1.
\ee
\noindent These two operators and the Hamiltonian ${\cal{H}}_R$ satisfy
the superalgebra. Therefore ${\cal{H}}_R$ is bounded from below and the
wavefunction given in (\ref{eq17.0}) is its exact groundstate for arbitrary
$\delta$. We have found a new, non-standard realization of supercharge
giving rise to a quantum mechanical super-Hamiltonian. We have used the
term non-standard in the sense that the explicit forms of $q$ and $q^{\dagger}$
involve higher order derivatives and also depends on the fermionic variables 
in a non-linear way.

The Hamiltonian ${\cal{H}}_R$ has dynamical $O(2,1) \times U(1)$ symmetry.
The radial excitations can be calculated easily by acting different
powers of the creation operator ${\cal{B}}_2^+$, defined in Appendix F
with ${\cal{B}}_0$ replaced by ${\cal{B}}_0 - \lambda \delta
\sum_i (x_i - x_{i+1})^{-2} \psi_i^{\dagger} \psi_i \psi_{i+1}^{\dagger}
\psi_{i+1}$, on $\phi_0$. Since
we have shown that ${\cal{H}}_R$ can be mapped to a set of free
superoscillators through the similarity transformation, the problem
now reduces to construct the eigenfunctions of ${\cal{H}}_R$ from
suitable free super-oscillator basis. We find the following $N_f=1$
solution of ${\cal{H}}_R$,
\be
\chi = T_1^{-1} \sum_i x_i^n \psi_i^{\dagger} |0>, \ \ n=0, 1, 2, 3.
\ee
\noindent We are not able to find any other physically acceptable
eigenstates.

Before ending this section, we just mention about the applicability of
this technique to other interacting fermionic theories. We recall at this
point that the Hubbard model can be viewed as an interacting
theory of two different XY models. Let us denote the Hamiltonian
in (\ref{eq6.3}) as $H^{\psi}$, where the superscript implies
that the fermionic variables in $H$ are denoted by $\psi=\{\psi_1, \psi_2,
\dots, \psi_N\}$. Now define a new Hamiltonian $\tilde{H}$ as,
\be
\tilde{H} = \frac{1}{2} \left ( H^{\psi} + H^{\tilde{\psi}} \right )
+ \delta \sum_i h_i \psi_i^{\dagger} \psi_i {\tilde{\psi}_i}^{\dagger}
\tilde{\psi}_i.
\label{hm}
\ee
\noindent We fix the convention that $\psi_i$ denotes the spin-up component,
while the $\tilde{\psi}_i$ denotes the spin-down component of the same
$i$th particle. With this identification in mind, it is easy to see that
the Hubbard model is contained in the Hamiltonian (\ref{hm}). In the freezing
limit, $W_i=0$, this Hubbard Hamiltonian can be decoupled from the parent
Hamiltonian. Due to the presence of the $O(2,1)$ symmetry in the rational case,
an infinite number of exact eigenstates corresponding to the radial
excitations can be calculated easily. Also following Ref. \cite{me},
$\tilde{H}$ can be mapped to a much simpler Hamiltonian
through a similarity transformation for the choice of rational potential.
In particular,
\bea
\tilde{H}_{sho} & = & H_{sho} +
\omega \sum_i \tilde{\psi}_i^{\dagger} \tilde{\psi}_i\nonumber \\
& = & \tilde{T}^{-1} \tilde{H} \tilde{T},
\eea
\noindent where $\tilde{T}= e^{\frac{\tilde{S}}{2}} e^W$. The $W$
in $\tilde{T}$  is given by (\ref{eq16}), while $\tilde{S}$ is determined by,
\be
\tilde{S} =  \frac{1}{2} \left ( S^{\psi} + S^{\tilde{\psi}} \right )
+ \delta \sum_i h_i \psi_i^{\dagger} \psi_i {\tilde{\psi}_i}^{\dagger}
\tilde{\psi}_i.
\ee
\noindent We would like to point out that one can in fact choose,
\be
G = \prod_{i=1}^N \prod_{k=1}^r (x_i - x_{i+k} )^{\lambda},
\ee
\noindent to produce an interacting fermionic theory with $r$th neighbor
interaction in the FL. A suitable deformation of such theories will
produce Hubbard model with $r$th neighbor interaction. However, finding
the exact eigenstates, other than those infinitely many radial excitations
due to the underlying $O(2,1)$ symmetry, is a nontrivial task and left
for a future investigation.

\section{Summary and discussions}

We have constructed and studied the supersymmetric version of a class of
many-particle Hamiltonians related to the short-range Dyson model. The
fermionic sector of these Hamiltonians contains inhomogeneous XY models
in an external magnetic field. We have shown that there exists a natural FL
for any super-model with classically stable minimum equilibrium configurations,
in which the fermionic degrees of freedom can be decoupled completely from the
 kinematical one. In the FL, the XY model decouples completely from the parent
super-model. This implies that the exact eigen-spectrum of the XY
spin chain can be constructed from the corresponding states of the full
super-model.

We have shown that the rational super-model for both the cases of
$A_{N+1}$ and $BC_{N+1}$ can be mapped to the set of
free superoscillators through the similarity transformation. This mapping
is very useful for constructing the exact eigenstates of the Hamiltonian,
which should be otherwise very difficult and intractable. We have constructed
many exact eigenstates of the super-model using this mapping.
For the $A_{N+1}$-type rational super-model, the spectrum is that of three 
free superoscillators with the frequencies $\omega, 2 \omega, 3 \omega$ and
one fermionic
oscillator with the frequency $4 \omega$. It appears that the spectrum is not
complete and also the disparity between the bosonic and the fermionic
excitations
in the spectrum is not clear at this point. The exact eigen-states
corresponding to this spectrum are obtained from the free superoscillator
basis which are invariant under the combined exchange of the bosonic and the
fermionic coordinates. Though there are no compelling reasons to choose only
such symmetric basis, our effort to construct a well-behaved eigen-function of
the super-model from a superoscillator basis without this permutation symmetry
has failed. 

For the $BC_{N+1}$-type super-model, the spectrum is that of a superoscillator
with the frequency $2 \omega$ and a fermionic oscillator with the frequency
$4 \omega$.
The disparity between the bosonic and the fermionic excitations are seen again
in this case as in its $A_{N+1}$ counterpart. It seems that the spectrum is
not complete. The exact eigenstates
corresponding to the exact spectrum are obtained from the free superoscillator
basis which are invariant under, (i) the combined exchange of the bosonic and
the fermionic coordinates and (ii) the simultaneous reflection of the bosonic
and the fermionic coordinates. Again, there are no compelling reasons to
consider only such superoscillator basis with the specified permutation and
reflection symmetry. Unfortunately, we are unable to find any
other basis which produces physically acceptable non-singular wavefunction
for the super-model.

We have also constructed the exact ground-state of the Trigonometric and the
Hyperbolic models. However, in absence of any mapping as in the
case of the rational model, the task of finding the excited states are
nontrivial and has not been addressed in this paper.

We have considered a suitable deformation of these super-models such that
an inhomogeneous XXZ spin chain can be obtained in the FL. This again opens
up an alternative method of studying the XXZ spin chains from super-models.
We are able to find the exact groundstate of this deformed super-model for 
the rational, trigonometric and the hyperbolic case. Finding the exact
ferromagnetic groundstate of a manifestly semi-positive Hamiltonian
will not surprise anybody. However, we have shown that the rational version
of these deformed super-models can be mapped to free superoscillators
through the similarity transformation. We have also constructed an exact
eigenstate of this rational model with the fermion number one. Unfortunately,
we are unable to find any more exact eigenstates of this model. However, we
believe that the mapping of the rational model to the free superoscillator
will allow us to obtain more exact states in future.

The supersymmetry of the deformed-super-model in terms of the usual realizations
of the supercharges is lost. However, for the rational case, this model is
mapped to the set of free superoscillators through the similarity
transformation. Using this mapping, we have shown that the
deformed-super-model with the rational potential is also supersymmetric in
terms of the new, non-standard realization of the supercharges. It remains to
be seen whether such realization of the supercharges in theories with
$O(2,1)$ symmetry has any further consequences or not.

It is known that the Haldane-Shastry spin chains\cite{hs} can be obtained from
the CMS model with internal degrees of freedom in the FL\cite{fl}. It is also
known that the super-CMS model describes a particular sector of the usual
CMS model with internal $SU(N)$ degrees of freedom\cite{bh,fm}. Thus, in the
FL, the super-CMS model describes only a particular sector of the
Haldane-Shastry spin
chains. On the other hand, the super-model considered in this paper
describes the full sector of different NN XY models in the FL. Further, the
deformed-super-models describe the full sector of different NN XXZ models
in the same limit. Unfortunately, at present, we are unable to ascertain the
exact solvability of these parent super-models or the deformed-super-models.
It might be recalled here that the super-potential $W$ with the constraints
(\ref{eq6.2}), (\ref{sup}) and (\ref{supsup}) produces (deformed-)super-models,
(i) which can be mapped to free superoscillators through the
similarity transformation and (ii) give rise to various spin chain
Hamiltonians in the FL. Even with all these constraints, there are many choices
for $W$. It would be nice if at least one $W$ can be chosen such that the
corresponding (deformed-)super-model is exactly solvable. 

\acknowledgements{We would like to thank Avinash Khare for a careful reading
of the manuscript and valuable comments. The work of PKG is supported by a
fellowship (P99231) of the Japan Society for the Promotion of Science.
This work is partially supported by the Grant-in-Aid for encouragement 
of Young Scientists (No. 12740231). }

\appendix{

\section{Jordan-Wigner transformation}

We reproduce the results concerning the Jordan-Wigner transformation in this
Appendix. The XXZ Hamiltonian is given by,
\be
H_{xxz} = \frac{1}{2} \sum_j \left ( \sigma_j^x \sigma_{j+1}^x +
\sigma_j^y \sigma_{j+1}^y \right ) +
\frac{\delta}{2} \sum_j \sigma_j^z \sigma_{j+1}^z,
\label{xxz}
\ee
\noindent where $\sigma$'s are the Pauli matrices and $\delta$ is the asymmetry
parameter. The XY model can be obtained by simply putting $\delta=0$. The above
Hamiltonian can be rewritten as,
\be
H_{xxz} = \sum_j \left [ \left ( \sigma_j^+ \sigma_{j+1}^- +
\sigma_j^- \sigma_{j+1}^+ \right )
+ \frac{\delta}{2} \sigma_j^z \sigma_{j+1}^z \right ], \ \
\sigma_j^{\pm} = \frac{1}{2} \left ( \sigma_j^x \pm i \sigma_j^y \right ).
\ee
\noindent We define the Jordan-Wigner transformation which transforms
the spin variables $\sigma_i$'s to the fermionic variables $\psi_i$'s,
\be
\psi_j = e^{i \pi \sum_{k=1}^{j-1} \sigma_k^+ \sigma_k^- } \sigma_j^-,\ \
\psi_j^{\dagger} = e^{- i \pi \sum_{k=1}^{j-1} \sigma_k^+ \sigma_k^-} 
\sigma_j^+ ,
\label{jw}
\ee
\noindent with,
\be
\sigma_i^+ \sigma_{i+1}^-= \psi_i^{\dagger} \psi_{i+1}, \ \
\sigma_i^- \sigma_{i+1}^+ = - \psi_i \psi_{i+1}^{\dagger}, \ \
\psi_i^{\dagger} \psi_i = \sigma_i^+ \sigma_i^- = \frac{1}{2} (1 + \sigma_i^z).
\label{id}
\ee
\noindent The inverse of (\ref{jw}) is given by,
\be
\sigma_j^- = e^{ i \pi \sum_{k=1}^{j-1} \psi_k^{\dagger} \psi_k} \psi_j, \ \
\sigma_j^+ = e^{ - i \pi \sum_{k=1}^{j-1} \psi_k^{\dagger} \psi_k}
\psi_j^{\dagger}.
\label{ijw}
\ee
\noindent Note that the fermionic variables satisfy the Clifford algebra
(\ref{eq1}). The Hamiltonian $H_{xxz}$ in (\ref{xxz}) with periodic
boundary condition now reads as,
\be
H_{xxz} = \sum_i \left ( \psi_i^{\dagger} \psi_{i+1} -
\psi_i \psi_{i+1}^{\dagger} \right ) + \frac{\delta}{2} \sum_i
n_i n_{i+1}.
\ee
\noindent We get different types of fermionic Hamiltonians in the FL of 
the super-models, which have similarity with the above Hamiltonian. The
use of the transformations (\ref{jw}), (\ref{id}) and (\ref{ijw}) will
enable us to write the corresponding spin Hamiltonian.

\section{Quadratic superpotential and XY model}

In this appendix, we discuss about supersymmetric model with the
superpotential being quadratic in the bosonic coordinates. The
bosonic and the fermionic degrees of freedom decouple automatically
for this kind of superpotentials. We further restrict the superpotential
to be of a particular form such that the fermionic part describes an
inhomogeneous XY model in an external non-uniform magnetic field. 
Let us choose the following form of the superpotential,
\be
W = \frac{1}{2} \sum_i a_i x_i^2 + \sum_i b_i x_i x_{i+1}, \ x_{N+i} = x_i,
\label{eq6.4}
\ee
\noindent where $a_i$'s and $b_i$'s are arbitrary constants. This amounts to
choosing $g_i=a_i$ and $h_i=b_i$ in equations (\ref{eq6.2}), (\ref{eq6.3}) and
(\ref{eq6.33}). The Hamiltonian $H_s$ (fermionic part of $H$) completely
decouples from the
super-Hamiltonian $H$ and describes the most general inhomogeneous isotropic XY
model in a non-uniform external magnetic field,
\be
H_s = \frac{1}{2} \sum_i \left [ a_i \sigma_i^z + b_i \left (
\sigma_i^x \sigma_{i+1}^x + \sigma_i^y \sigma_{i+1}^y \right ) \right ].
\ee
\noindent The XY model as
the fermionic part of a supersymmetric quantum mechanics for the particular
choice of the superpotential (\ref{eq6.4}) with $a_i=a$ and $b_i=b$ for all $i$
has first been noticed in \cite{ritten}. Later, the same model has been studied
from the point of view of supersymmetry breaking in quantum mechanical
models \cite{jk} with finite degrees of freedom.

The techniques of supersymmetry are not much useful in studying the spectrum
of $H_s$, since it is completely decoupled from the super-Hamiltonian $H$.
Diagonalising the Hamiltonian $H$ essentially implies the diagonalisation of
$H_k$ ( bosonic part of $H$) and $H_s$ separately or vice versa. However, we
notice an interesting relation
between the supersymmetric phase of $H$ and the ferromagnetic phase of $H_s$.
In particular, the groundstate of $H$ with zero eigen-value is given by,
\be
\phi_0 = e^{- ( \frac{1}{2} \sum_i a_i x_i^2 + \sum_i b_i x_i x_{i+1} ) } |0>.
\label{eq6.5}
\ee
\noindent The fermionic vacuum $|0>$ in $\phi_0$ describes the completely
ferromagnetic state in terms of spin degrees of freedom. The normalizability
of $\phi_0$ demands,
\be
b_i^2 \leq a_i a_{i+1}, \ i=1, 2, \dots N-1, \ N > 5,
\label{eq6.6}
\ee
\noindent with each $a_i$ being positive.
There are additional relations involving $b_N$ for $N < 5$. Thus,
$\phi_0$ with the constraints (\ref{eq6.6}) is the supersymmetric groundstate
of $H$. The supersymmetry is broken when the relations (\ref{eq6.6}) are
not satisfied, and the groundstate energy of the Hamiltonian $H$ will be
positive definite with some well-behaved eigenfunctions. We now notice that
the groundstate of $H_s$ is given by the fermionic vacuum or the completely
ferromagnetic state, only in the case when the relation (\ref{eq6.6}) is
satisfied.
Thus, there is a one to one correspondence between the supersymmetric phase of
$H$ and the ferromagnetic phase of $H_s$. We now discuss a particular example
of a homogeneous isotropic XY model in an external staggered magnetic field
by choosing,
\be
a_i = m + (-1)^i m_s, \ \ b_i = b .
\label{eq6.7}
\ee
\noindent We find from Eq. (\ref{eq6.6}) that the supersymmetric phase and,
hence, the ferromagnetic phase is described in the region $ m > (b^2 +
m_s^2)^{\frac{1}{2}}$ of the parameter space. It is known\cite{alc} that the
anti-ferromagnetic to ferromagnetic transition of $H_s$, with the choice
of $a_i$ and $b_i$ given by (\ref{eq6.7}), occurs exactly at this critical
value.

\section{Classical Equilibrium configurations}
The classical equilibrium configuration of $H$ is determined by Eq.
(\ref{eq17.1}). In terms of rescaled variables $x_i \rightarrow 
\sqrt{\omega} x_i$, this reads,
\be
x_i + \left [ (x_{i-1} - x_i )^{-1} - (x_i - x_{i+1})^{-1} \right ] = 0.
\label{A.1}
\ee
\noindent It is trivial to solve the above equations for $N=2$, giving
$x_1=-1, x_2=1$ or vice-versa. For N=3, the zeroes of the third Hermite
polynomial satisfy the above equations. We have the following solution for
$N=4$,
\be
x_1=0, \ x_2=-\sqrt{2},\ x_3=0, \ x_4=\sqrt{2}.
\ee
\noindent Note that although the coordinates $x_1$ and $x_3$ are identical,
it does not produces any singularity because
the interaction depends on the distance between two NN particles only.
We do not know any solution for $N \geq 5$. The equation (\ref{A.1}) for
arbitrary $N$ can be rewritten as,
\be
\eta_{i-1} \eta_{i+1} (\eta_i^2 -2) +
\eta_i ( \eta_{i-1} + \eta_{i+1} ) = 0, \ \
\eta_i = x_i - x_{i+1}.
\label{A.2}
\ee
\noindent Now note that $\eta_i=\eta$ for all $i$ is not a solution of the
above equation. Thus, the equilibrium configurations do not constitute an
uniform lattice. In passing, we also note the following identities for the
equilibrium configuration,
\be
\sum_i x_i = 0, \ \sum_i x_i^2 = N, \ \sum_i x_i^3 = 0.
\label{A.3}
\ee
\noindent Though it would be nice to obtain such identities for arbitrary
powers, as in the case of usual CMS, we are unable to find them.

The classical equilibrium configuration of the $BC_{N+1}$-type rational
model in terms of the rescaled variables
$x_i \rightarrow \sqrt{\frac{\omega}{\lambda}} x_i$
reads as,
\be
x_i + 2 x_i \left [ \frac{1}{x_{i-1}^2 - x_{i}^2} - \frac{1}{x_i^2 -
x_{i+1}^2} \right ] - \frac{\lambda_1}{\lambda x_i} = 0.
\ee
\noindent One can check again that the uniform lattice is not a solution
of this equation. Note the identity,
\be
\sum_i x_i^2 = \left (2 + \frac{\lambda_1}{\lambda} \right ) N,
\ee
\noindent for the equilibrium configurations. 

\section{Trigonometric model}

We choose the superpotential as,
\be
W = - ln \prod_i sin^{\beta} \left [ \frac{\pi}{L} ( x_i - x_{i+1} ) \right ],
\ x_{N+i}= x_i.
\label{t1}
\ee
\noindent This produces the following expressions for $g_i$ and $h_i$,
\be
h_i = - \frac{\beta \pi}{L} sin^{-2} [\frac{\pi}{L} (x_{i} - x_{i+1})], \ \
g_i = - (h_i + h_{i-1}),
\ee
\noindent and the Hamiltonian $H=H_k + H_s$ now reads as,
\bea
H_s & = & \frac{\beta \pi^2}{2 L^2} \sum_i  sin^{-2} q_i
\left ( n_i + n_{i+1}  - 2 \left ( \psi_i^{\dagger}
\psi_{i+1} - \psi_i \psi_{i+1}^{\dagger} \right ) \right ),\nonumber \\
H_k & = & \frac{1}{2} \sum_i p_i^2
+ \frac{\beta^2 \pi^2}{ L^2} \sum_i \left ( sin^{-2} \ q_i
 -  cot q_{i-1} \ ccot q_i -1 \right ), \ \
q_i= \frac{\pi}{L} (x_i - x_{i+1}). 
\eea
\noindent The supersymmetric groundstate of $H$ for $\beta > 0$ is obtained as,
\be
\psi_0^+ = \prod_i sin^{\beta} q_i |0>.
\ee
\noindent On the other hand, for $\beta < 0$, the supersymmetric groundstate
is given by,
\be
\psi_0^- = \prod_i sin^{-\beta} |\bar{0}>.
\ee
\noindent The groundstate $\psi_0^+$ is in the zero fermion sector, while it is
in the $N$ fermion sector for $\psi_0^-$. It is amusing to note, unlike in the
rational case, the supersymmetric phase is obtained for $\beta > 0$ as well as
for $\beta < 0$. The Trigonometric model is defined on a circle and there is
no need to add extra confining potential to stabilize the system. This
difference with the rational case is manifested in determining the
supersymmetry-preserving and supersymmetry-breaking phases of the two models.

Consider a new Hamiltonian ${\cal{H}}$,
\be
{\cal{H}}= H + \frac{\beta \pi^2 \delta }{2 L^2} \sum _i
sin^{-2} q_i \ \left ( 1 - n_i n_{i+1} \right ).
\ee
\noindent Both $\psi_0^{+}$ and $\psi_0^{-}$ are valid eigenstates of
${\cal{H}}$ with zero eigenvalue for $\beta > 0$ and $\beta < 0$, respectively.
The super-Hamiltonian $H$ is bounded from below, $H \geq 0$, due to the
supersymmetry. Now note the following identity for each index $i$,
\be
( 1 - n_i n_{i+1} ) = \frac{1}{2} ( n_i - n_{i+1} )^2 \geq 0.
\ee
\noindent As a consequence, the Hamiltonian ${\cal{H}}$ is also bounded from
below for (i) $\beta > 0, \delta > 0$ and (ii) $\beta < 0, \delta < 0$. Thus,
both $\psi_0^+$ and $\psi_0^-$ are in fact the groundstates of ${\cal{H}}$
in the region (i) and (ii) of the parameter space, respectively.

In the freezing limit, the XY model and the XXZ model can be
obtained from $H$ and ${\cal{H}}$, respectively. In this
limit, the bosonic coordinates in $H_s$ and ${\cal{H}}_s$ are determined
by the equation,
\be
W_i = \frac{\beta \pi}{L} \left [ cot q_{i-1} - cot q_i \right ] =0.
\ee
\noindent This equation has the solution that all the particles are
equispaced on a circle with the angular separation between two particles
given by, $\frac{2 \pi}{N}$. Thus, the corresponding spin chain Hamiltonians
are homogeneous XY and XXZ models in an external uniform magnetic field. These
models are in fact exactly solvable. The ground state of the XY model is
massless and located at the boundary of the ferromagnetic gapped phase.
On the other hand, the ground state of the XXZ model is massive and it
is in the ferromagnetic gapped phase.

\section{Hyperbolic model}

We choose the superpotential as,
\be
W = - ln \prod_i sinh^{\beta} \left (  x_i - x_{i+1}  \right )
+ \gamma \sum_i e^{2 x_i}, \ x_{N+i}= x_i.
\label{h1}
\ee
\noindent The first term of (\ref{h1}) can be obtained from (\ref{t1}) by
rescaling, $ x_j \rightarrow \frac{L}{\pi} i x_j$, where $i$ is the imaginary
number. The second term has been introduced to stabilize the system. It
produces a confining Morse potential. The expression for $g_i$ and $h_i$ are,
\be
h_i = \beta sinh^{-2} q_i, \ \ g_i = - (h_i + h_{i-1}) + 4 \gamma
e^{2 x_i}, \ \ q_i=x_i - x_{i+1},
\ee
\noindent with the Hamiltonian $H=H_k + H_s$ given by,
\bea
H_k & = & \sum_i \left [ \frac{p_i^2}{2} + \beta^2 \left (sinh^{-2} q_i
- coth q_{i-1} coth q_i + 1 \right ) + 2 \gamma^2 e^{4 x_i}
- 4 \gamma \beta e^{2 x_i} \right ],\nonumber \\
H_s & = & \frac{\beta}{2} \sum_i \left [ - sinh^{-2} q_i \left \{ n_i
+ n_{i+1} - 2 \left (\psi_i^{\dagger} \psi_{i+1} -
\psi _i \psi_{i+1}^{\dagger} \right )\right \} +
\frac{4 \gamma}{\beta} e^{2 x_i} n_i \right ]. 
\eea
\noindent  Note that the bosonic sub-model of $H$ ( i.e. the zero-fermion
sector ) is the nearest-neighbor variant of the corresponding CMS
Hamiltonian \cite{fra}.
The ground-state of $H$ in the supersymmetry-preserving phase ($\beta > 0,
\gamma > 0$ ) is given by,
\be
\psi_0 = \prod_i sinh^{\beta} q_i e^{- \gamma \sum_i e^{2 x_i}} |0>.
\ee
\noindent This is also the groundstate of,
\be
{\cal{H}} = H + \frac{\beta \delta^2}{2} \sum_i sinh^{-2} q_i \ 
\psi_i^{\dagger} \psi_i \psi_{i+1}^{\dagger} \psi_{i+1},
\ee
\noindent with zero eigenvalue. We do not know the groundstates of $H$ and
${\cal{H}}$ in other regions of the parameter space of $\beta$ and $\gamma$.
This is because the `method of duality' used for the rational case does
not work for the specific confining potential of this Hyperbolic model.
In the FL, the Hamiltonian $H$ and ${\cal{H}}$ gives rise to XY and XXZ
Hamiltonian, respectively. In this limit, the bosonic coordinates in $H_s$
and ${\cal{H}}_s$ are determined by the following relation,
\be
W_i = \beta \left ( coth p_{i-1} - cothp_i \right ) + 2 \gamma
e^{2 x_i} = 0.
\ee
\noindent We do not know the solution of this equation.

\section{$OSP(2|2)$ super-algebra}

We show in this appendix that the Hamiltonian $H$ in (\ref{eq6.1}),
with the superpotential given by Eq. (\ref{sup}) and (\ref{supsup}),
has dynamical $OSp(2|2)$ supersymmetry. We work with $\omega=1$ in this
Appendix without loss of any generality. We first define the following four
supercharges,
\bea
&& q= \frac{1}{2} \sum_{i=1}^N \psi_i^{\dagger} ( p_i + i w_i - i x_i), \ \
q^{\dagger}= \frac{1}{2} \sum_{i=1}^N \psi_i ( p_i - i w_i + i x_i),\nonumber \\
&& {\tilde{q}} = \frac{1}{2} \sum_{i=1}^N \psi_i \left ( p_i - i w_i -
i x_i \right ), \ \
{\tilde{q}}^{\dagger} = \frac{1}{2} \sum_{i=1}^N \psi_i^{\dagger}
\left ( p_i + i w_i + i x_i \right ).
\label{a0}
\eea
\noindent The Hamiltonian $H$ is given in terms of $q$ and
$q^{\dagger}$as, $H = 2 \{q, q^{\dagger}\}$. The dual Hamiltonian
${\cal{H}}^d$ can be constructed in terms of $\tilde{q}$ and
${\tilde{q}}^{\dagger}$, ${\cal{H}}^d= 2 \{{\tilde{q}},
{\tilde{q}}^{\dagger}\}$. We define the following operators,
\bea
&& h = \frac{1}{2} \left ( H + {\cal{H}}^d \right ), \ \
U = \frac{1}{2}  \left ( H - {\cal{H}}^d \right ),\ \
E = \sum_i x_i \frac{\partial}{\partial x_i},\nonumber \\
&& {\cal{B}}_2 ^- = {\cal{B}}_0 -\frac{1}{4} \sum_i x_i^2
- \frac{1}{4} \left ( N + 2 E \right ), \ \
{\cal{B}}_2 ^+ = {\cal{B}}_0 - \frac{1}{4} \sum_i x_i^2
+ \frac{1}{4} \left ( N + 2 E \right ),\nonumber \\
&& {\cal{B}}_0 = \frac{1}{4} \sum_i \left ( p_i^2 + w_i^2 + w_{ii} -
2 \sum_{j} w_{ij} \psi_i^{\dagger} \psi_j \right ).
\label{a1}
\eea
\noindent The bosonic operators ${\cal{B}}_2^{\pm}$ and $h$ satisfies
the following relations,
\be
[h, {\cal{B}}_2^{\pm}] = \pm 2 {\cal{B}}_2^{\pm}, \ \
[{\cal{B}}_2^-, {\cal{B}}_2^+] = h.
\label{a2}
\ee
\noindent The commutator relation $[E,{\cal{B}}_0]=- 2 {\cal{B}}_0$
has been used in deriving the above equations. The $U(1)$ generator $U$
commutes with ${\cal{B}}_2^{\pm}$ and $h$.

The non-vanishing anticommutators among $q$, $q^{\dagger}$, $\tilde{q}$
and $\tilde{q}^{\dagger}$ are,
\be
\{q, q^{\dagger}\} = \frac{1}{2} ( h + U ) , \ \
\{\tilde{q}, \tilde{q}^{\dagger}\} = \frac{1}{2} ( h - U ), \ \
\{q^{\dagger}, \tilde{q}^{\dagger}\} = {\cal{B}}_2^{+}, \ \
\{q, \tilde{q}\} = {\cal{B}}_2^{-} . \ \
\label{a3}
\ee
\noindent Observe that the relation,
\be
H = {\cal{H}}_d + 2 U,
\label{a4}
\ee
\noindent which is useful in determining the spectrum in the supersymmetry 
breaking phase, follows easily from the first two equations of (\ref{a3}).

The other non-vanishing commutators are,
\bea
&& [{\cal{B}}_2^+, q] = - \tilde{q}^{\dagger}, \ \
[{\cal{B}}_2^+, \tilde{q}] = - q^{\dagger}, \ \
[{\cal{B}}_2^-, \tilde{q}^{\dagger}] = q, \ \
[{\cal{B}}_2^-, q^{\dagger}] = \tilde{q},\nonumber \\
&& [h, q^{\dagger}]= q^{\dagger}, \ \
[h, q]= - q, \ \
[h, \tilde{q}]=-\tilde{q}, \ \
[h,\tilde{q}^{\dagger}] = \tilde{q}^{\dagger},\nonumber \\
&& [U, \tilde{q}] = - \tilde{q}, \ \
[U, \tilde{q}^{\dagger}] = \tilde{q}^{\dagger}, \ \
[U, q^{\dagger}] = - q^{\dagger}, \ \
[U, q] = q.
\label{a5}
\eea}

\end{document}